\documentclass[12pt]{article}
\usepackage{graphicx}
\begin{document}
\centerline{\bf Species Numbers versus Area in Chowdhury Ecosystems}
\bigskip

Dietrich Stauffer* and Andrzej P\c{e}kalski**
\bigskip

Department of Genetics, Institute of Microbiology, University of Wroc{\l}aw, 
ul. Przybyszewskiego 63/77, PL-54148 Wroc{\l}aw, Poland;
visiting from Institute for Theoretical Physics, Cologne University, 
D-50923 K\"oln, Euroland. e-mail: stauffer@thp.uni-koeln.de 

**Institute of Theoretical Physics, University of Wroc{\l}aw, pl. M.Borna 9,
PL-51-204 Wroc{\l}aw, Poland. e-mail: apekal@ift.uni.wroc.pl

\bigskip
Abstract: For a fixed number of food levels on a lattice, the Chowdhury
model for ecosystem is simulated and found to give a number of different
species which approaches exponentially a plateau if the considered area
is increased.
\bigskip

\begin{figure}[hbt]
\begin{center}
\includegraphics[angle=-90,scale=0.5]{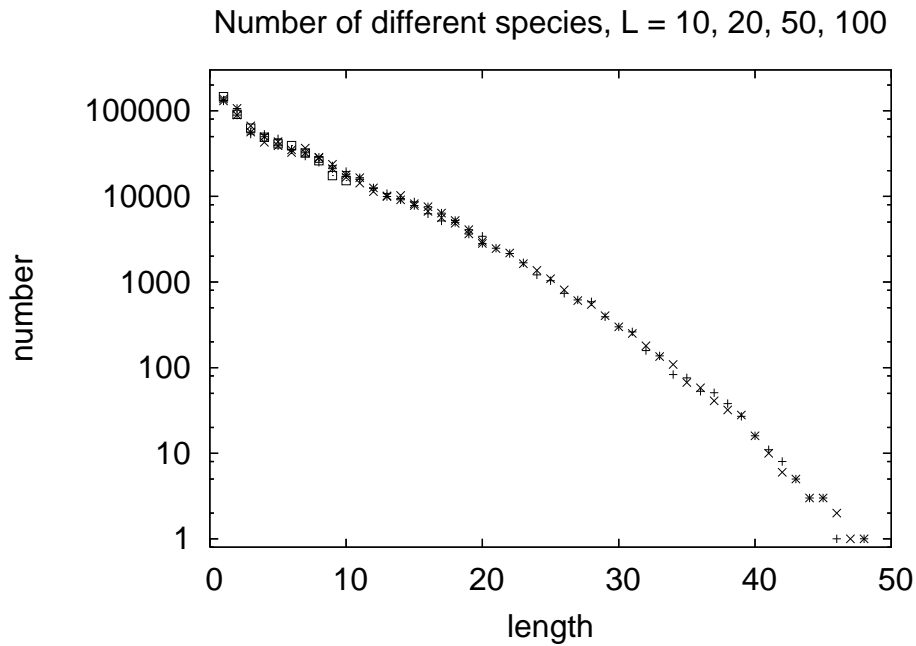}
\end{center}
\caption{Number of new species in $R$-squares which are not yet found 
in the inner squares of size $R-1$, summed over 20,000 time steps (iterations)
for lattices of linear dimension $L$ = 10 (+), 20 (x), 50 (stars), and 100 
(squares) from bottom to top (one sample each,
except sum over ten samples for $L = 10$). The counted numbers are divided by
the number $2R-1$ of involved lattice sites and rounded down to integer 
values. Each lattice carried six food levels with 1 to 32 ecological niches.}
\end{figure}

\begin{figure}[hbt]
\begin{center}
\includegraphics[angle=-90,scale=0.40]{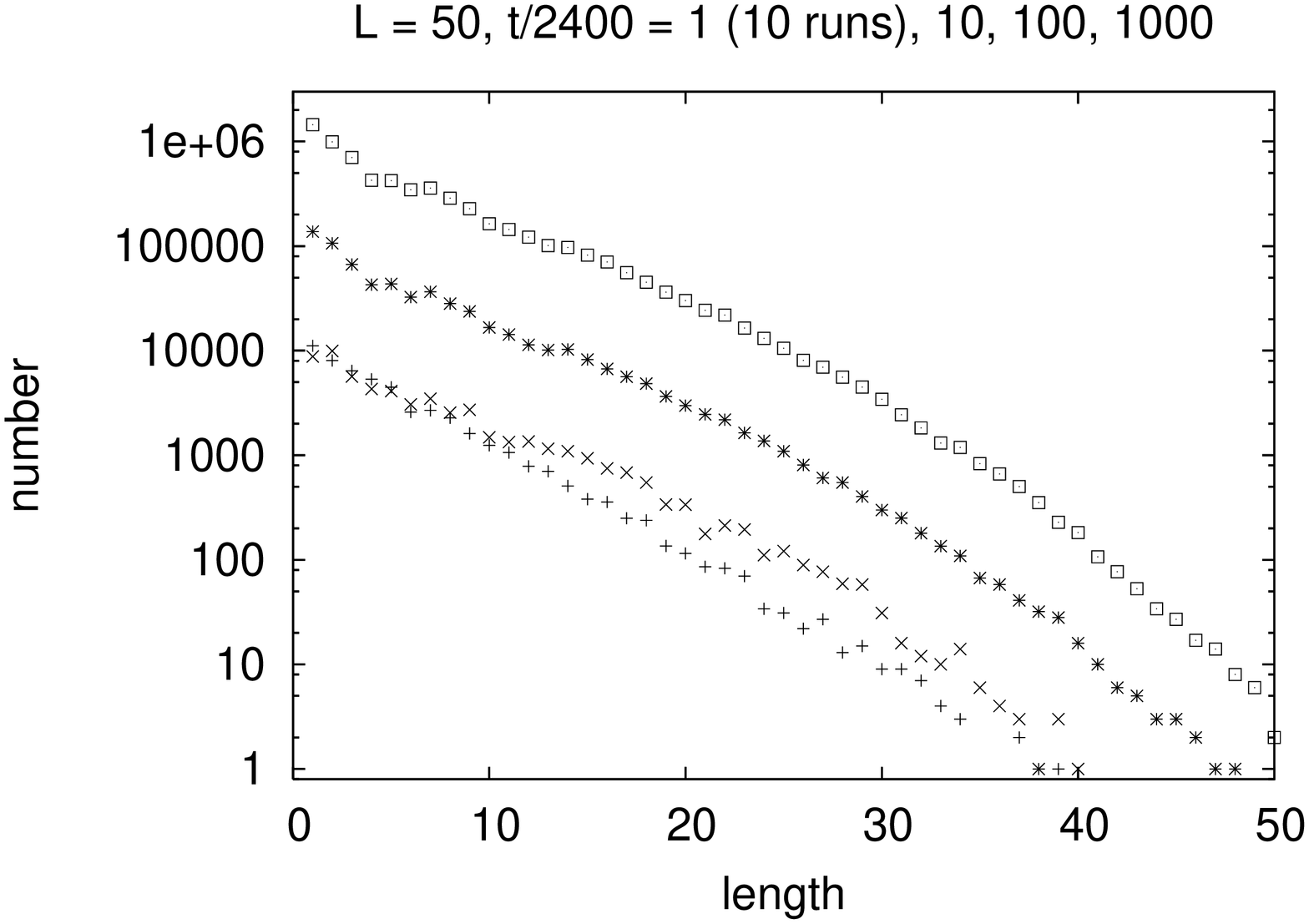}
\includegraphics[angle=-90,scale=0.40]{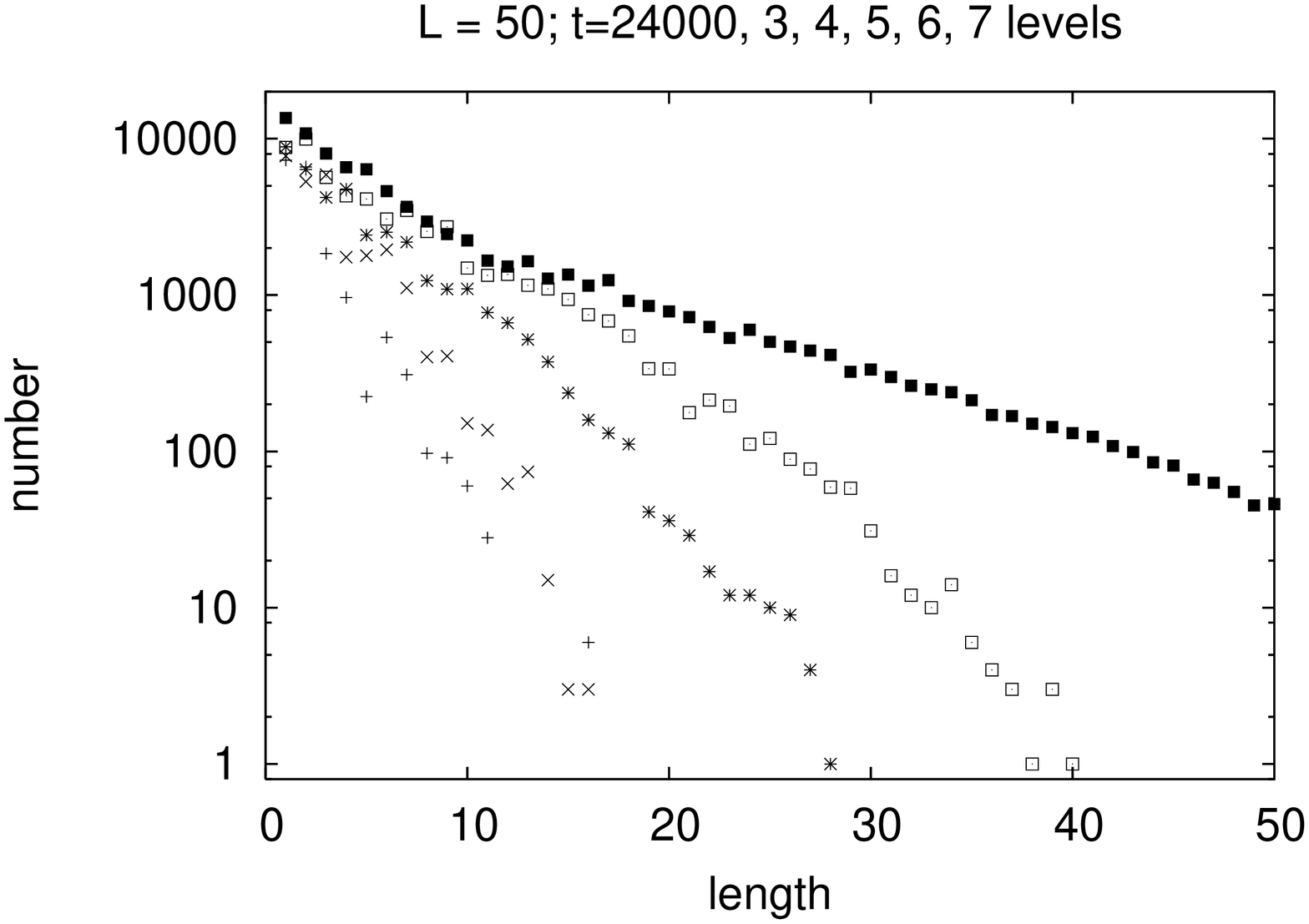}
\end{center}
\caption{As Fig.1 but in part a for t = 2,400 to 2,400,000 iterations at 
$L = 50$, and in part b for three to seven food levels (from bottom to 
top) at $L = 50, \; t = 24,000$.}
\end{figure}
The Chowdhury model \cite{debch} is one of several individual-based models for 
ecosystems with many (prey-predator) species \cite{droz,veit}. We use it here
with the parameters of \cite{rohde}, that means a low speciation rate 0.001,
and low diffusivity 0.001, and a fixed number (usually six) of trophic 
layers. We refer to \cite{debch,rohde,newbook} for more details on this model.
Here we check if the number of different species which are found in a square
of size $R \times R$ increases with a small power of the area $R^2$, as 
summarized in \cite{brazil}.

So after some equilibration time (one sixth of the total number of iterations)
we sum at each iteration
over all different species found in a corner of size $R \times R$ of
the $L \times L$ lattice, with $1 \le R \le L$ and $10 \le L \le 100$. To see
better the trends with increasing $R$ we look at the $R$-derivative of this
number of species, that means we determine how many new species were found
in a square of side length $R$ which were not already found in its inner square
of side length $R-1$. A species in this asexual model is defined to consist
of all animals occupying the corresponding ecological niches on the considered
lattice sites. 

Figure 1 shows that this number of new species decays roughly exponentially
with increasing $R$ (though for small $R$ also a power law fits). Thus the
total number of species approaches exponentially a plateau for large $R$, in 
contrast to the desired power law \cite{brazil}. This exp($-R/\xi$) decay with 
$R$ 
is found to be independent of the simulation time $t$, Fig.2a, while the decay
length $\xi$ increases when the number of food levels increases from 3 to 7,
Fig.2b. An inhomogeneous ecosystem with different but fixed numbers of food
levels thus might give a power law as a superposition of many exponential laws.

We thank PapaSmorf Cebrat for hospitality during our collaboration, anf the
Wroc{\l}aw-K\"oln university partnership for partial support.

\end{document}